\documentclass[aps,prd,twocolumn,amsmath,amssymb,amsfonts,nofootinbib,superscriptaddress,altaffilletter]{revtex4-2}

\usepackage{graphicx}
\usepackage{hyperref}
\hypersetup{
  bookmarksopen=true
}
\usepackage{ulem}
\usepackage{amssymb}
\usepackage{amsmath}
\usepackage{color}
\usepackage{supertabular}
\usepackage{dcolumn}
\usepackage[mathscr]{eucal}
\usepackage{mathrsfs}
\usepackage{siunitx}

\def\sci#1#2{#1\times10^{#2}}

\def\RAJ{\textrm{RA}_{\textrm J2000}}
\def\DECJ{\textrm{DEC}_{\textrm J2000}}

\begin{document}

\title{Expanded atlas of the sky in continuous gravitational waves}

\author{Vladimir Dergachev}
\email{vladimir.dergachev@aei.mpg.de}
\affiliation{Max Planck Institute for Gravitational Physics (Albert Einstein Institute), Callinstrasse 38, 30167 Hannover, Germany}

\author{Maria Alessandra Papa}
\email{maria.alessandra.papa@aei.mpg.de}
\affiliation{Max Planck Institute for Gravitational Physics (Albert Einstein Institute), Callinstrasse 38, 30167 Hannover, Germany}
\affiliation{Leibniz Universit\"at Hannover, D-30167 Hannover, Germany}

\begin{abstract}
We present the full release of the atlas of continuous gravitational waves, covering frequencies from 20\,Hz to 1700\,Hz and spindowns from $\sci{-5}{-10}$ to $\sci{5}{-10}$\,Hz/s. Compared to the early atlas release, we have extended the frequency range and have performed follow-up on the outliers. Conducting continuous wave searches is computationally intensive and time-consuming. The atlas facilitates the execution of new searches with relatively minimal computing resources.
\end{abstract}

\maketitle

\section{Introduction}

Continuous gravitational waves are very weak nearly monochromatic signals that are significantly more challenging to detect than the coalescences of binary black holes or neutron stars.

One expects continuous waves from rapidly rotating neutron stars exhibiting a sustained non-axisymmetric deformation $\varepsilon$. The intrinsic amplitude of the gravitational wave signal at a distance $d$ from the source is proportional to the deformation, increases with the square of the signal frequency $f$ and decreases with the inverse of the distance to the source $d$:
\begin{equation}
h_0=\frac{4\pi^2G I_{zz} f^2 \varepsilon}{c^4 d},
\label{eq:epsilon}
\end{equation}
with $I_{zz}=10^{38}$\,kg\,m$^2$ being the moment of inertia of the star with respect to the principal axis aligned with the rotation axis. 

Gravitational waves carry away energy, which slows-down the pulsar rotation.
Observations of  pulsars provide data on the spin and spin-down rates of neutron stars, suggesting that their deformations are very small. This is further supported by the non-detections of increasingly sensitive surveys \cite{O3aDataSet,EatHO3a,lvc_O3_allsky,lvc_O3_allsky2} and population studies \cite{Pagliaro:2023bvi}. We hence have developed the Falcon pipeline 
optimized for weak signals coming from nearby neutron stars with ellipticities \cite{loosely_coherent, loosely_coherent2, loosely_coherent3, allsky3} more than two orders of magnitude below the maximum estimates \cite{crust_limit}. The resulting searches can detect the weakest signals probed by all-sky searches.

\begin{figure}[htbp]
\includegraphics[width=3.3in]{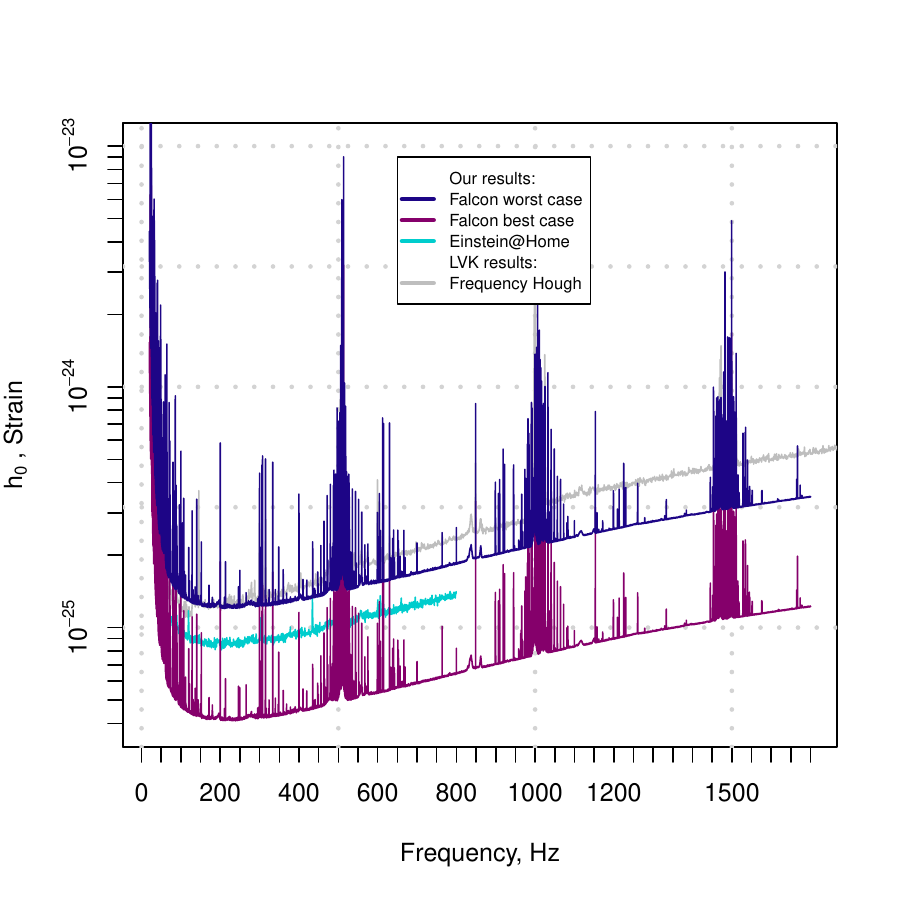}
\caption[Upper limits]{
\label{fig:amplitudeULs}
Gravitational wave intrinsic amplitude $h_0$ upper limits at 95\% confidence as a function of signal frequency. The upper limits measure the sensitivity of the search. We also plot latest LIGO/Virgo and Einstein@Home all-sky population average results \cite{lvc_O3_allsky2, EatHO3a,o3a_atlas2}.
}
\end{figure}

Since all-sky searches for continuous gravitational waves like ours are very computationally intensive, we have started releasing large-scale data from our Falcon surveys for others to use in their own analyses, e.g. to follow-up, to cross-correlate with other astrophysical probes or for coincidences with their own gravitational wave search results on other data sets. We call such releases ``the atlas" \cite{o3a_atlas1}, as they comprehensively cover the sky in many searched frequency bands.  In \cite{o3a_atlas2} we provide an example of atlas usage by constraining gravitational emission from the supernova remnants Vela Jr and G189.1+3.0. 

Our last release \cite{o3a_atlas2} was from an all-sky search on LIGO O3a data \cite{O3aDataSet} for signals with frequency up to 1500 Hz. We shared that atlas even before completing the follow-up of the most interesting results. Here we report results of the follow-up of atlas hot spots, extended to 1700\,Hz and a cross-check against the O3b data \cite{O3bDataSet}. 509 outliers pass our follow-up pipeline using O3a data, and of those only 22 survive the cross-check against O3b data. Of those 22 outliers, one is attributed to a detector artifact, while the rest are due to simulated signals injected during the O3 science run.

The results presented here cover signal frequencies up to 1700 Hz, representing a 50\% increase in the searched parameter space volume with respect to \cite{o3a_atlas2}, as the number of templates scales with the cube of the frequency.

A new release of the atlas accompanies these results. The atlas was constructed from stage 1 ($\approx 87$\%) and stage 2 ($\approx 13$\%) analysis results. Thus the majority of points are waiting to be more deeply explored.

The atlas includes signal-to-noise ratio (SNR) data and upper limits on the intrinsic gravitational wave amplitude $h_0$ as a function of sky position and frequency. The upper limits are provided for both worst-case wave polarization and best-case (circular) polarization. Polarization-specific upper limits can be computed for arbitrary inclination angle $\iota$ and polarization angle $\psi$ using Eq.s (20) and (21) in \cite{functional_upper_limits} with the parameter values $c_{1-14}$ provided in the atlas. Together with the atlas we also provide example code for polarization-specific upper limits.

Our upper limits have been established across the entire frequency band with no exclusions. The upper limits maximized over sky-position are shown in Figure \ref{fig:amplitudeULs}.

The atlas has been constructed in such a way, that to lookup data for a potential signal at a given frequency and coming from a given direction in the sky one only needs to inspect results in the 45 mHz atlas-frequency-band containing the signal frequency and from the atlas sky position nearest to that direction in the sky (using distance in the celestial sphere).

\section{The search}

Our search used O3a data for the main analysis, with cross-checks performed against O3b data. The setup was previously described in \cite{o3a_atlas2}. The extension to 1700\,Hz was performed without any modifications. As before, the upper limits were derived using the universal statistics \cite{universal_statistics}, which infers the absence of signals with amplitude above a certain level based on the level of power in the underlying data. The universal statistics is designed to be valid for  arbitrarily distributed data and hence we do not need compute-intensive Monte Carlos to establish upper limits. We have however performed Monte Carlo campaigns testing the signal-recovery capabilities of our pipeline on simulated signals for validation purposes.
\begin{table}[htbp]
\begin{center}
\begin{tabular}{rD{.}{.}{2}D{.}{.}{3}}\hline
Stage & \multicolumn{1}{c}{Coherence length (days)} & \multicolumn{1}{c}{Minimum SNR}\\
\hline
\hline
1  & 0.5 & 6 \\
2  & 1 & 8 \\
3  & 2 & 9 \\
4  & 6 & 16 \\
\hline
\end{tabular}
\end{center}
\caption{Parameters for each stage of the search. Stage 4 refines outlier parameters by using denser sampling of the parameter space and then subjects them to an additional consistency check by comparing outlier parameters from analyses of individual interferometer data. Only the first two stages were used to construct the atlas, while subsequent stages were used for outlier analysis.}
\label{tab:pipeline_parameters}
\end{table}
 
The outlier follow-up is new to this paper and used three stages, as shown in Table \ref{tab:pipeline_parameters}. Through stages 2-4, the coherence length increases, and the size of the region covered by a loosely coherent template decreases, resulting in refined outlier parameters. At every stage only candidates with SNR above a given threshold are passed to the next stage. The thresholds are deliberately chosen to be low in order to reduce the likelihood of losing detections due to overly aggressive SNR cutoff.

Each outlier from stage 4 is further checked for consistency across the different detector combinations, 
by searching the data from the different detectors separately and in coherent combination, all using the same stage 4 setup.
We require outlier parameters from these different searches to be consistent, and discard outliers with frequency mismatch  $> 2$\,\textmu Hz or with spindown mismatch $> 0.3$\,pHz/s. The outliers that pass this check are included with this paper \cite{data}.

\begin{figure}[htbp]
\includegraphics[width=3.3in]{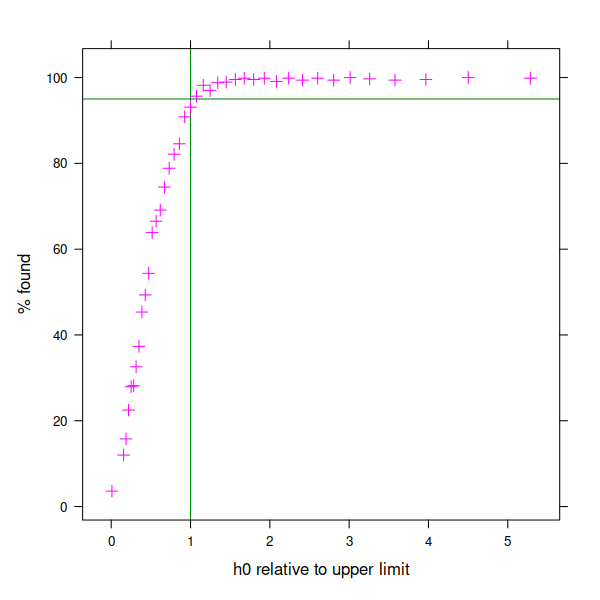}
\caption[Injection recovery]{
\label{fig:injection_recovery}
Injection recovery versus injection strain normalized to the upper limit established on the same data without injection. The injections were performed in 500-2000\,Hz band and included regions highly contaminated with detector artifacts.
}
\end{figure}

The cross-check with O3b data was performed last, using a very simple requirement that the outlier frequency be within $10$\,\textmu Hz between O3a and O3b datasets. The O3b searches also used the stage 4 setup. Our simulations show that this results in recovery of at least 95\% of injections with strength at/above the upper limit value (Figure  \ref{fig:injection_recovery}).

\begin{table*}[htbp]
\begin{center}
\begin{tabular}{D{.}{.}{3}D{.}{.}{5}D{.}{.}{1}D{.}{.}{3}D{.}{.}{3}D{.}{.}{2}D{.}{.}{2}D{.}{.}{4}D{.}{.}{4}D{.}{.}{4}D{.}{.}{4}}\hline
\multicolumn{1}{c}{SNR}   &  \multicolumn{1}{c}{$f$} & \multicolumn{1}{c}{$\dot{f}$} &  \multicolumn{1}{c}{$\RAJ$}  & \multicolumn{1}{c}{$\DECJ$} & \multicolumn{1}{c}{iota} & \multicolumn{1}{c}{psi} & \multicolumn{1}{c}{$f_{\textrm{O3a}}-f_{\textrm{O3b}}$}  & \multicolumn{1}{c}{$f_{\textrm{O3a}}-f_{\textrm{inj}}$} & \multicolumn{1}{c}{$\dot{f}_{\textrm{O3a}}-\dot{f}_{\textrm{inj}}$} & \multicolumn{1}{c}{dist$_\textrm{inj}$}\\
\multicolumn{1}{c}{(stage 4)}	&  \multicolumn{1}{c}{Hz}	&  \multicolumn{1}{c}{pHz/s} & \multicolumn{1}{c}{degrees} & \multicolumn{1}{c}{degrees} & \multicolumn{1}{c}{degrees} & \multicolumn{1}{c}{degrees} & \multicolumn{1}{c}{$\mu$\,Hz} & \multicolumn{1}{c}{$\mu$\,Hz} & \multicolumn{1}{c}{pHz/s}& \multicolumn{1}{c}{arcsec} \\
\hline \hline
 55.6 & 20.50003 & 0.0 & 269.7440 & 66.5787 & 135.00 & 80.00 & -8.2 &  &  & \\
 439.8 & 52.80832 & 0.2 & 302.6456 & -83.8261 & 56.25 & 72.00 & -5.3 & -0.7 & 0.2 & 47.4\\
 440.5 & 52.80832 & -0.0 & 302.6316 & -83.8374 & 56.25 & 72.00 & 3.4 & -0.2 & 0.0 & 6.4\\
 445.4 & 52.80832 & 0.0 & 302.6428 & -83.8390 & 56.25 & 72.00 & -0.5 & -0.2 & 0.0 & 6.2\\
 433.3 & 52.80832 & -0.0 & 302.6646 & -83.8368 & 56.25 & 72.00 & -2.4 & -0.2 & 0.0 & 17.0\\
 435.7 & 52.80832 & 0.2 & 302.5930 & -83.8244 & 56.25 & 72.00 & -0.5 & -0.2 & 0.2 & 54.7\\
 436.0 & 52.80832 & 0.0 & 302.6642 & -83.8378 & 56.25 & 72.00 & 0.5 & -0.2 & 0.0 & 15.3\\
 435.7 & 52.80832 & 0.2 & 302.5929 & -83.8244 & 56.25 & 72.00 & -0.5 & -0.2 & 0.2 & 54.7\\
 436.6 & 52.80832 & -0.0 & 302.6570 & -83.8370 & 56.25 & 72.00 & 0.5 & -0.2 & 0.0 & 14.0\\
 434.2 & 52.80832 & -0.4 & 302.7008 & -83.8673 & 56.25 & 72.00 & 5.8 & 0.3 & -0.4 & 105.3\\
 433.5 & 52.80832 & -0.4 & 302.6881 & -83.8642 & 56.25 & 72.00 & 5.8 & 0.3 & -0.4 & 93.2\\
 441.6 & 52.80832 & -0.4 & 302.6930 & -83.8651 & 56.25 & 72.00 & 5.8 & 0.3 & -0.4 & 96.7\\
 55.4 & 108.85716 & 0.0 & 178.3665 & -33.4376 & 101.25 & 30.00 & 0.5 & -0.3 & 0.0 & 18.6\\
 55.4 & 108.85716 & -0.0 & 178.3750 & -33.4342 & 101.25 & 30.00 & -3.4 & 0.1 & 0.0 & 11.5\\
 82.4 & 265.57505 & -4.4 & 71.5535 & -56.2150 & 45.00 & 50.00 & 4.3 & 0.0 & -0.2 & 9.5\\
 82.2 & 265.57505 & -4.0 & 71.5510 & -56.2180 & 45.00 & 50.00 & -2.4 & 0.0 & 0.2 & 2.7\\
 84.9 & 265.57505 & -4.0 & 71.5512 & -56.2180 & 45.00 & 50.00 & -2.4 & 0.0 & 0.2 & 2.5\\
 85.3 & 265.57505 & -3.8 & 71.5509 & -56.2220 & 45.00 & 50.00 & -5.8 & 0.5 & 0.3 & 16.2\\
 100.1 & 575.16350 & -0.4 & 215.2574 & 3.4481 & 135.00 & 60.00 & 6.3 & -1.2 & -0.3 & 15.6\\
 98.7 & 575.16351 & 0.2 & 215.2575 & 3.4416 & 135.00 & 70.00 & -6.8 & -0.3 & 0.3 & 9.7\\
 110.7 & 763.84732 & 0.4 & 198.8926 & 75.6906 & 123.75 & 9.00 & -1.4 & -1.2 & 0.4 & 7.3\\
 1055.5 & 848.93498 & -300.0 & 37.3940 & -29.4525 & 67.50 & 22.50 & 0.5 & 0.1 & 0.0 & 0.6\\
\hline
\end{tabular}
\caption[Outliers produced by the detection pipeline]{Outliers produced by the detection pipeline that passed the coincidence check with O3b data. All outliers except the one at 20.5\,Hz are due to simulated signals ``hardware-injected'' during the science run for validation purposes. Their parameters are listed in Table \ref{tab:injections}. The last three columns show differences between outliers and injection frequency, frequency derivative, and location on the sky.
All signal frequencies, for both the O3a and the O3b analyses, refer to GPS epoch $1246070000$ (2019 Jul 2 02:33:02 UTC), which is in the middle of the O3a period. Outliers stemming from the same injection present the same values of iota and psi, because these lie on a grid, and exact coincidences are expected. Louder injections produce more outliers, and some of them are associated with signal waveforms quite farther away from the injection parameters than the closest match. Lower frequency signals are generally less well localized in the sky, as expected since the sky resolution increases with frequency.
}
\label{tab:Outliers}
\end{center}
\end{table*}

\begin{table}[htbp]
\begin{center}
\begin{tabular}{l D{.}{.}{3.5} r c r r r r r}
\hline
 Label & \multicolumn{1}{c}{$f$} & \multicolumn{1}{c}{$\dot{f}$} &  SNR & UL/$h_0$ & $\Delta f$ & In & Found\\
& \multicolumn{1}{c}{Hz} & Hz/s & (atlas) & \% & mHz & \\
\hline
\hline
ip0 & 265.57505 & -4.15e-12 & 28.5 & 122.5 & -0.1 & Yes & Yes\\
ip1 & 848.93498 & -3e-10  & 393.0 & 119.9 & -0.1 & Yes & Yes\\
ip2 & 575.16351 & -1.37e-13  & 39.3 & 138.5 & 0.0 & Yes & Yes\\
ip3 & 108.85716 & -1.46e-17  & 23.7 & 141.6 & 0.1 & Yes & Yes\\
ip4 & 1390.60583 & -2.54e-08  & 7.6 & 21.3 & -7.7 & No & No \\
ip5 & 52.80832 & -4.03e-18  & 155.9 & 130.2 & 0.0 & Yes & Yes \\
ip6 & 145.39178 & -6.73e-09  & 8.4 & 25.0 & -11.2 & No & No\\
ip7 & 1220.42586 & -1.12e-09  & 7.3 & 68.1 & 3.6 & No & No\\
ip8 & 190.03185 & -8.65e-09  & 8.9 & 83.8 & -2.9 & No & No\\
ip9 & 763.84732 & -1.45e-17  & 39.1 & 135.1 & 0.1 & Yes & Yes\\
ip10 & 26.33210 & -8.5e-11  & 63.9 & 124.9 & 0.0 & Yes & No \\
ip11 & 31.42470 & -5.07e-13  & 93.2 & 400.9 & -12.1 & Yes & No\\
ip12 & 37.75581 & -6.25e-09  & 14.0 & 156.5 & 4.0 & No & No\\
ip16* & 234.56700 & 0  & 8.3 & 29.6 & 42.7 & No & No\\
ip17* & 890.12300 & 0  & 8.1 & 103.6 & 23.6 & No & No \\
\hline
\end{tabular}

\caption[Hardware injections]{This table shows the frequency and sky position parameters of the hardware-injected continuous wave signals in the data and how they appear in the atlas data.  We show our upper limit UL at the injection parameters, and how it compares with the injection amplitude $h_0$: if the upper limits are correct, UL/$h_0$ has to be $> 100$\%.
$\Delta f$ is the difference between the frequency at the SNR peak in the atlas and the injection frequency. We show all the hardware injections within 20-1700\,Hz range, including those outside of our search space, as indicated by the ``In'' column -- for  instance ip16 and ip17 are signals from neutron stars in binary systems.  We use the reference time (GPS epoch) $t_0=1246070000$ (2019 Jul 2 02:33:02 UTC). The script {\tt spatial\_index\_example2.R} provided with the atlas shows how to obtain this data from the atlas. }
\label{tab:injections}
\end{center}
\end{table}

The full list of outliers that passed the O3b cross-check is shown in Table \ref{tab:Outliers}. All except one correspond to one of hardware-injected simulated signals (Table \ref{tab:injections}).

The outlier at 20.5\,Hz is associated with a known low-frequency comb of instrumental artifacts \cite{line_catalog}.

Out of eight hardware-injected signals within our parameter space, we detected six. The two missed signals, ip10 and ip11, are located in a highly contaminated low-frequency region.

The injection ip10 produces intermediate outliers in both O3a and O3b data but is rejected by the coincidence check because the H1 interferometer lacks the sensitivity to detect it.

The injection ip11 is a factor of 4 below the upper limit and does not progress to the stage 4 follow-up. Given the large number of candidates in the first stages and the low sky resolution in the low frequency range of this injection, it is hard to say whether some candidate from the earlier stages was associated to that injection.

\section{Conclusions}
We have performed a wide-band search of O3a data covering frequencies from 20 to 1700\,Hz. No significant outliers have passed the cross-check with O3b data during follow-up. The high frequency extension allows us to expand the best-case reach of our search to $250$\,pc for $\varepsilon > 10^{-8}$. We make the atlas of upper limits, signal-to-noise ratios and associated data available for further study. The atlas also includes several examples, written in {\tt R}, that can be a starting point for future searches.

\begin{acknowledgments}
The authors thank the scientists, engineers and technicians of LIGO, whose hard work and dedication produced the data that made this search possible.

The search was performed on the ATLAS cluster at AEI Hannover. We thank Bruce Allen, Carsten Aulbert and Henning Fehrmann for their support.

This research has made use of data or software obtained from the Gravitational Wave Open Science Center (gw-openscience.org), a service of LIGO Laboratory, the LIGO Scientific Collaboration, the Virgo Collaboration, and KAGRA. 
\end{acknowledgments}

\end{document}